\documentclass[journal]{IEEEtran}
\usepackage{cite}

\usepackage{graphicx}

\begin{document}
\bibliographystyle{plainnat}
\newcommand{\cm}[1]{\ensuremath{ {{\rm cm}^{#1}}}}
\newcommand{\ntp}[2]{\ensuremath{#1\times10^{#2} } }
\newcommand{\asb}[2]{\ensuremath{#1_{\rm #2} }}

\begin{titlepage}

\thispagestyle{empty}
\def\thefootnote{\fnsymbol{footnote}}       

\begin{center}
\mbox{ }

\end{center}\begin{center}
\vskip 1.0cm
{\Huge\bf
Measurements of Charge Transfer Inefficiency in
}
\vspace{2mm}

{\Huge\bf
a~CCD~with~High-Speed~Column\,Parallel\,Readout
}
\vskip 1cm
{\LARGE
Andr\'e Sopczak$^1$,
Khaled Bekhouche$^1$,  
Chris Damerell$^2$, \\
Tim Greenshaw$^3$, 
Michal Koziel$^1$,  
Konstantin Stefanov$^2$,\\ 
\smallskip
Tuomo Tikkanen$^3$,
Tim Woolliscroft$^3$, 
Steve Worm$^2$
\bigskip}

{\Large
$^1$Lancaster University, UK\\
$^2$STFC Rutherford Appleton Laboratory, UK\\
\smallskip
$^3$Liverpool University, UK
}

\vskip 2.5cm
\centerline{\Large \bf Abstract}
\end{center}

\vskip 2.cm
\hspace*{-0.5cm}
\begin{picture}(0.001,0.001)(0,0)
\put(,0){
\begin{minipage}{\textwidth}
\Large
\renewcommand{\baselinestretch} {1.2}
Charge Coupled Devices (CCDs) have been successfully used in several high energy physics experiments over the past two decades. 
Their high spatial resolution and thin sensitive layers make them an excellent tool for studying short-lived particles. 
The Linear Collider Flavour Identification (LCFI) collaboration is developing Column-Parallel CCDs (CPCCDs) for the vertex 
detector of a future Linear Collider. The CPCCDs can be read out many times faster than standard CCDs, 
significantly increasing their operating speed. A test stand for measuring the charge transfer inefficiency (CTI) of a 
prototype CPCCD has been set up. Studies of the CTI have been performed at a range of readout frequencies and operating temperatures.

\normalsize
\vspace{3.5cm}
\begin{center}
{\sl \large
Presented on behalf of the LCFI Collaboration at the \\
IEEE 2008 Nuclear Science Symposium, Dresden, Germany
\vspace{-6cm}
}
\end{center}
\end{minipage}
}
\end{picture}
\vfill

\end{titlepage}

\newpage
\thispagestyle{empty}
\mbox{ }
\newpage
\setcounter{page}{0}
\title{Measurements of Charge Transfer Inefficiency in a CCD with
       High-Speed Column Parallel Readout}

\author{Andr\'e~Sopczak,~\IEEEmembership{Member,~IEEE,}\thanks{A. Sopczak is with Lancaster University, UK. 
Presented on behalf of the}\thanks{~~LCFI Collaboration; E-mail: andre.sopczak@cern.ch}
Khaled~Bekhouche,\thanks{K. Bekhouche is with Lancaster University, UK}
Chris~Damerell,\thanks{C. Damerell is with STFC Rutherford Appleton Laboratory, UK}
Tim~Greenshaw,\thanks{T. Greenshaw is with Liverpool University, UK}
Michal~Koziel,\thanks{M. Koziel is with Lancaster University, UK}
Konstantin~Stefanov,\thanks{K. Stefanov is with STFC Rutherford Appleton Laboratory, UK}
Tuomo~Tikkanen,\thanks{T. Tikkanen is with Liverpool University, UK (currently with Leicester}\thanks{~~University, UK)}
Tim Woolliscroft,\thanks{T. Woolliscroft is with Liverpool University, UK}
Steve Worm\thanks{S. Worm is with STFC Rutherford Appleton Laboratory, UK}}
\maketitle

\begin{abstract}
Charge Coupled Devices (CCDs) have been successfully used in several high energy physics experiments over the past two decades. 
Their high spatial resolution and thin sensitive layers make them an excellent tool for studying short-lived particles. 
The Linear Collider Flavour Identification (LCFI) collaboration is developing Column-Parallel CCDs (CPCCDs) for the vertex 
detector of a future Linear Collider. The CPCCDs can be read out many times faster than standard CCDs, 
significantly increasing their operating speed. A test stand for measuring the charge transfer inefficiency (CTI) of a 
prototype CPCCD has been set up. Studies of the CTI have been performed at a range of readout frequencies and operating temperatures.
\end{abstract}

\section{Introduction}
The study of radiation hardness is crucial for the application of CCD detectors in High Energy Physics 
experiments~\cite{Damerell:1997vv,Stefanov,LCFI_web}. The LCFI collaboration has been developing and testing new CCD detectors
 for about 10 years~\cite{Damerell:1997vv,Stefanov,LCFI_web,Worm,Greenshaw}. 
Previous experimental results on CCD radiation hardness were reported for example in~\cite{Marconi,Brau2000,Brau2005}. 
Several models increased the understanding of radiation damage effects in CCDs~\cite{Mohsen,Hopkins,Hardy}. 
The measurements and analyses reported in this paper have been carried out in the LCFI collaboration~\cite{LCFI_web}. 
Simulation and modelling of CCD radiation hardness effects for a CCD prototype with sequential readout was reported at 
IEEE'2005~\cite{Sopczak2005}, comparing full TCAD simulations with analytic models was reported at IEEE'2006~\cite{Sopczak2006} 
and in Ref.~\cite{Sopczak2007}, simulation and modelling of a CCD  prototype with column parallel readout (CPCCD) at 
IEEE'2007~\cite{IEEE_Honolulu} and in Ref.~\cite{Sopczak2008}.

This work focuses on experimental  measurements and a method to 
determine the charge transfer inefficiency (CTI) performed with the CPCCD at a test stand at Liverpool University. 

\begin{figure}
    \centering
    \includegraphics[width=0.8\columnwidth,clip]{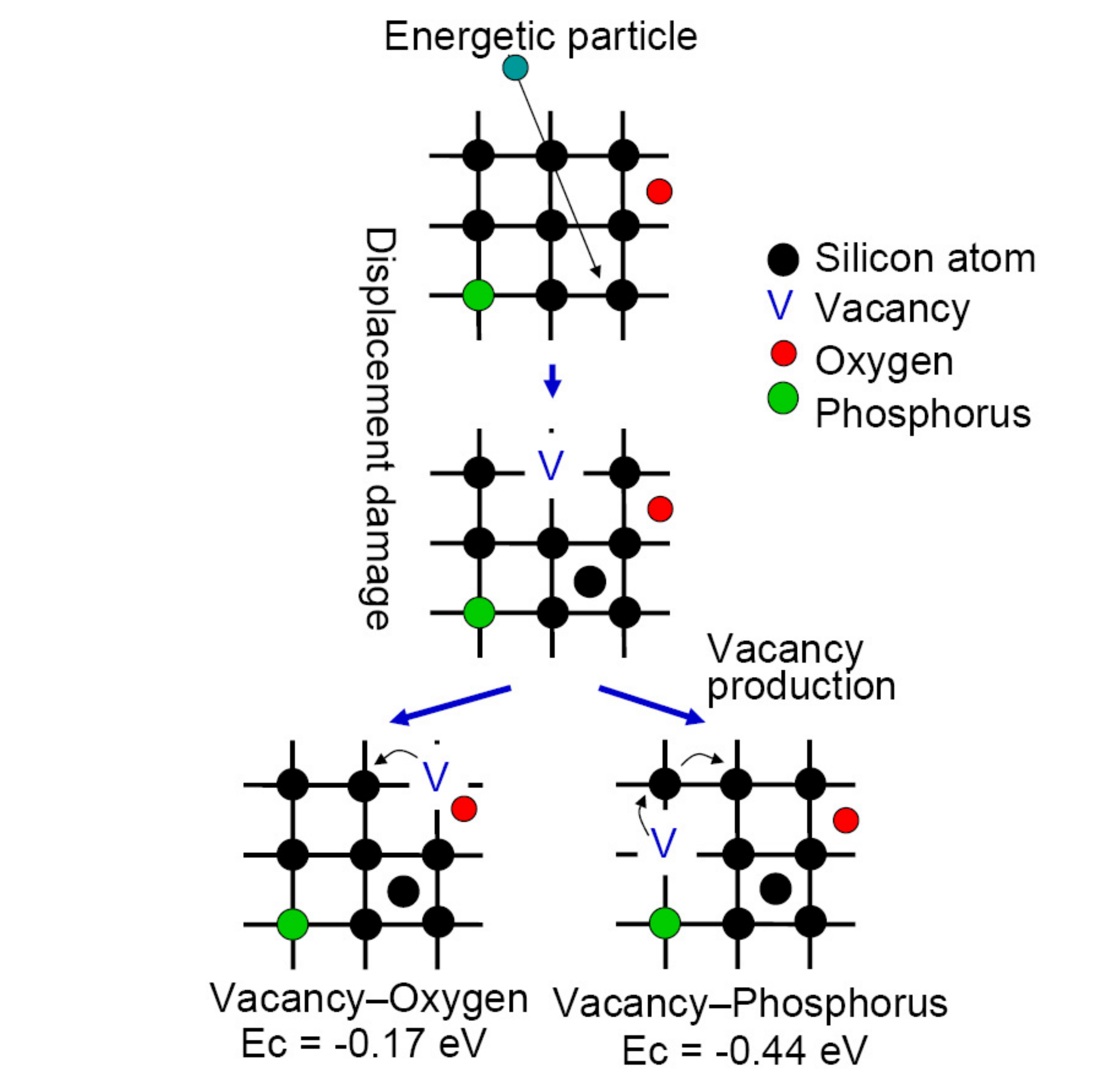}
\vspace*{-3mm}
    \caption{Schematic view of how radiation damage is created in phosphorus-doped silicon. 
              Two types of vacancy defect are created: 
             Vacancy-Oxygen (electron trap at $0.17$~eV below
     the conduction band) and Vacancy-Phosphorus (electron trap at $0.44$~eV below the conduction band).}
\vspace*{-5mm}
    \label{fig:Radiation_Damage}
\end{figure}

The environment with high radiation near the interaction point at a future Linear Collider creates damage to the CCD material 
which leads to defects acting as electron traps in the silicon, as indicated in Fig.~\ref{fig:Radiation_Damage}. 
The mechanism of creating traps has been discussed in the literature, for example in
 Refs.~\cite{Walker,Saks,Srour}. These traps result in charge transfer inefficiency. 
In a phosphorus-doped device, two types of traps are created. The first one is relatively shallow with energy level $0.17$~eV below the  conduction band and the 
second is deep with energy level $0.44$~eV below the conduction band. These traps result in charge transfer inefficiency. 

The column parallel technology is in development to cope with the required
readout rate. CPC1 is a two-phase CCD prototype capable of $50$~MHz readout frequency. 
In this paper we demonstrate the method to determine the CTI value with an un-irradiated CPCCD (CPC1).
\section{Test stand for CCD operation and cryostat unit}
\begin{figure}
    \centering
    \includegraphics[width=0.7\columnwidth,clip]{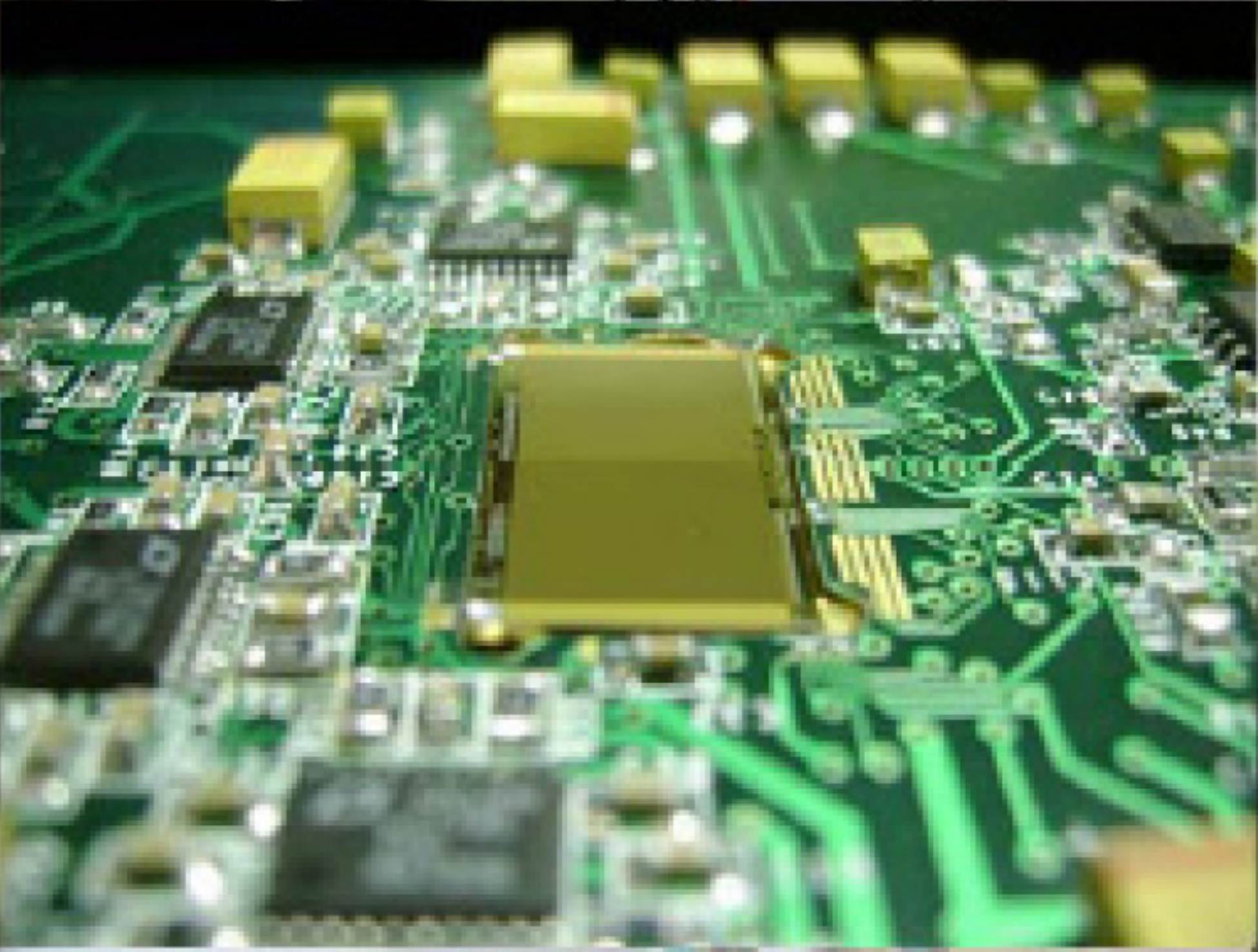}
    \caption{CPC1 (in the center of the picture) with external electronics.}
    \label{fig:CPCCD_Electronics_Photo}
\vspace*{-3mm}
\end{figure}

A test stand has been set up with readout electronics and a cryostat unit. The temperature range of the cryostat is
 from room temperature  down to about $-140~^\circ$C. This temperature has been achieved with cold nitrogen gas by boiling liquid nitrogen. 
The very low operating temperature is required to obtain sensitivity to the CTI peak structure for $0.17$~eV traps 
where the peak position is expected to be near $-130~^\circ$C~\cite{Sopczak2008}. 
Initial measurements have been performed on an un-irradiated device in standalone mode, where the signal from four columns of the CCD 
were amplified and connected to external ADCs. An $^{55}$Fe source emitting $6$~keV X-rays was attached to a holder 
at a distance of $5$~cm from the CCD to provide the signal charge.
Figure~\ref{fig:CPCCD_Electronics_Photo} shows a picture of CPC1 and the associated control and 
front-end readout electronics, which were also placed in the cryostat, wherefrom the four outputs were fed 
to rack-mounted amplifiers and ADCs.

It was observed that the CTI increased sharply when the amplitude of the sinusoidal clock pulses applied to read out the 
CCD was reduced to less than $2$~V peak-to-peak. In order to suppress any effect from the clock amplitude, settings for each 
data acquisition were tuned to produce $3.00~\rm V_{pp}$ clock pulses. 
The performance of the drivers limited the maximum clocking frequency to about $22$~MHz. 
The settings depended on the frequency and had to be adjusted for 
each cryostat temperature.
\section{Signal measurement and removing charge sharing by using a 3$\times$3 cluster method}
\begin{figure}
    \centering
    \includegraphics[width=\columnwidth,clip]{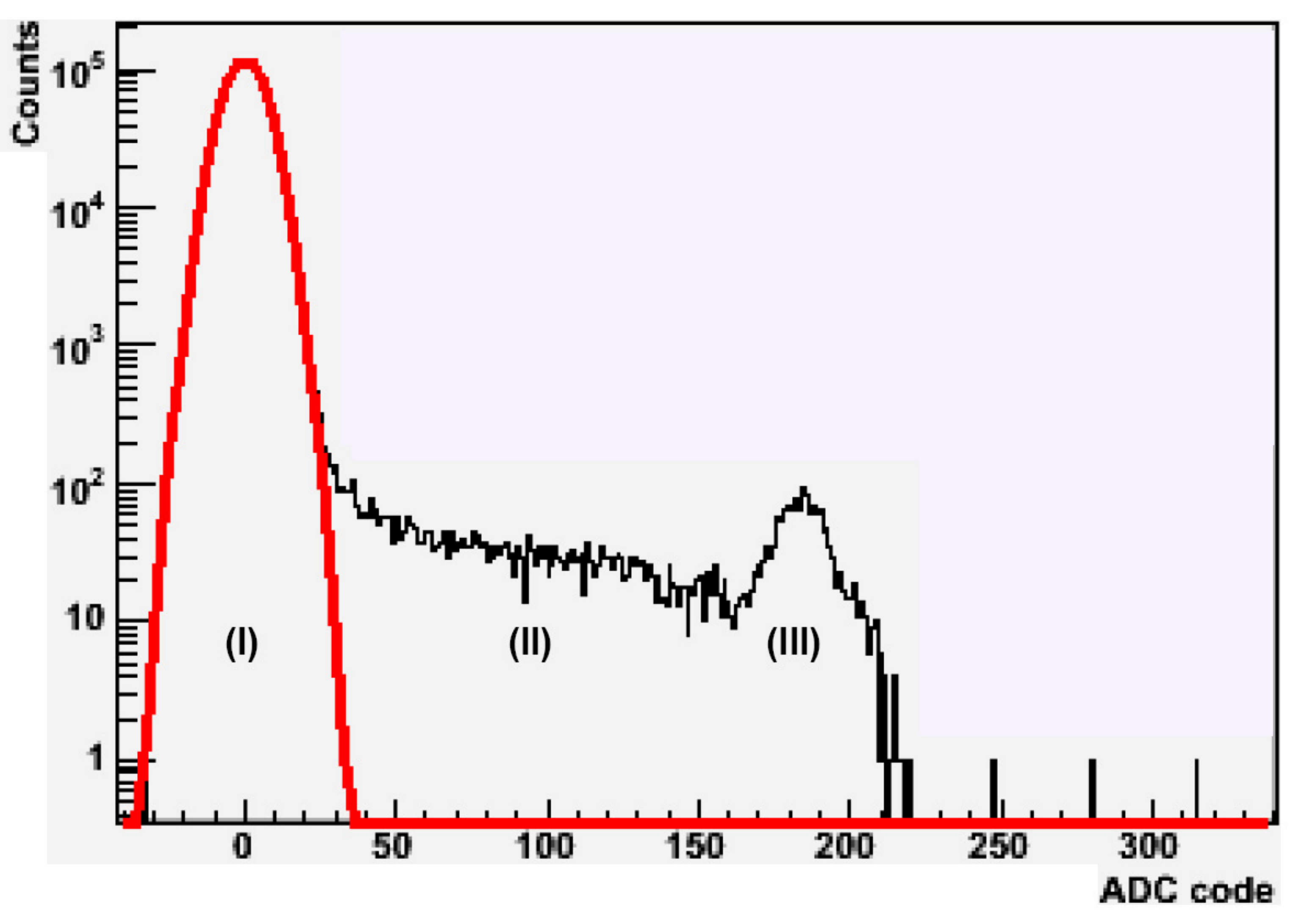}
    \caption{Distribution of ADC codes for channel (column) two. Three regions are observed:
    (I) the high peak region which represents the noise,
    (II) the region separating the two peaks which represents the charge sharing between pixels, and 
    (III) the X-ray peak region which represents the collected charge in a single pixel.
    The noise peak is fitted by a Gaussian function to determine the noise threshold. 
    The data was taken at $-80~^\circ$C with $8$~MHz readout frequency.}
    \label{fig:Fit_NoisePeak}
\vspace*{-2mm}
\end{figure}

The fast ADCs convert the signal charge after amplification with a wideband preamplifier. 
The four columns are read out in 4 channels by ADCs.
Three columns (channels 1 to 3) are adjacent whereas channel 4 is separate. 
The signal charges of 400 pixels  plus 10 overclocks were 
acquired in 3000--5000 frames per measurement. 
This leads to an accurate statistical precision. The collected  data has been 
analysed using the ROOT package~\cite{Root_Web}. 
First, we begin by applying correlated double sampling, where the difference between the signals of 
two consecutive pixels is taken to be the signal charge collected by the latter pixel. 
This decreases electronic noise. Next, we remove common mode noise components for each column
with the help of the overclocks. Overclocks are charges read out after the last pixel of the column. We average all overclocks of 
a column and subtract the average from the data of all pixels in the column.
As an example for channel 2 Fig.~\ref{fig:Fit_NoisePeak} shows the distribution of ADC codes for all recorded pixels
and all frames of one measurement.
The analysis is performed on column 2 because it is situated 
between columns 1 and 3. This allows to apply a 3$\times$3 cluster method.

Figure~\ref{fig:Fit_NoisePeak} illustrates that there are 3 regions in ADC codes. The first region (I) has a high
 number of pixels with low ADC codes from electronics noise only (no charge from X-rays). 
The noise is fitted by a Gaussian function in order to determine the noise threshold. The charge-sharing
 region (II) represents the signal charge shared between 2 or more pixels. 
Region (III) is  the signal region where the charge is collected in a single pixel that will be analysed to determine the CTI. 

\begin{figure}
    \centering
    \includegraphics[width=\columnwidth,clip]{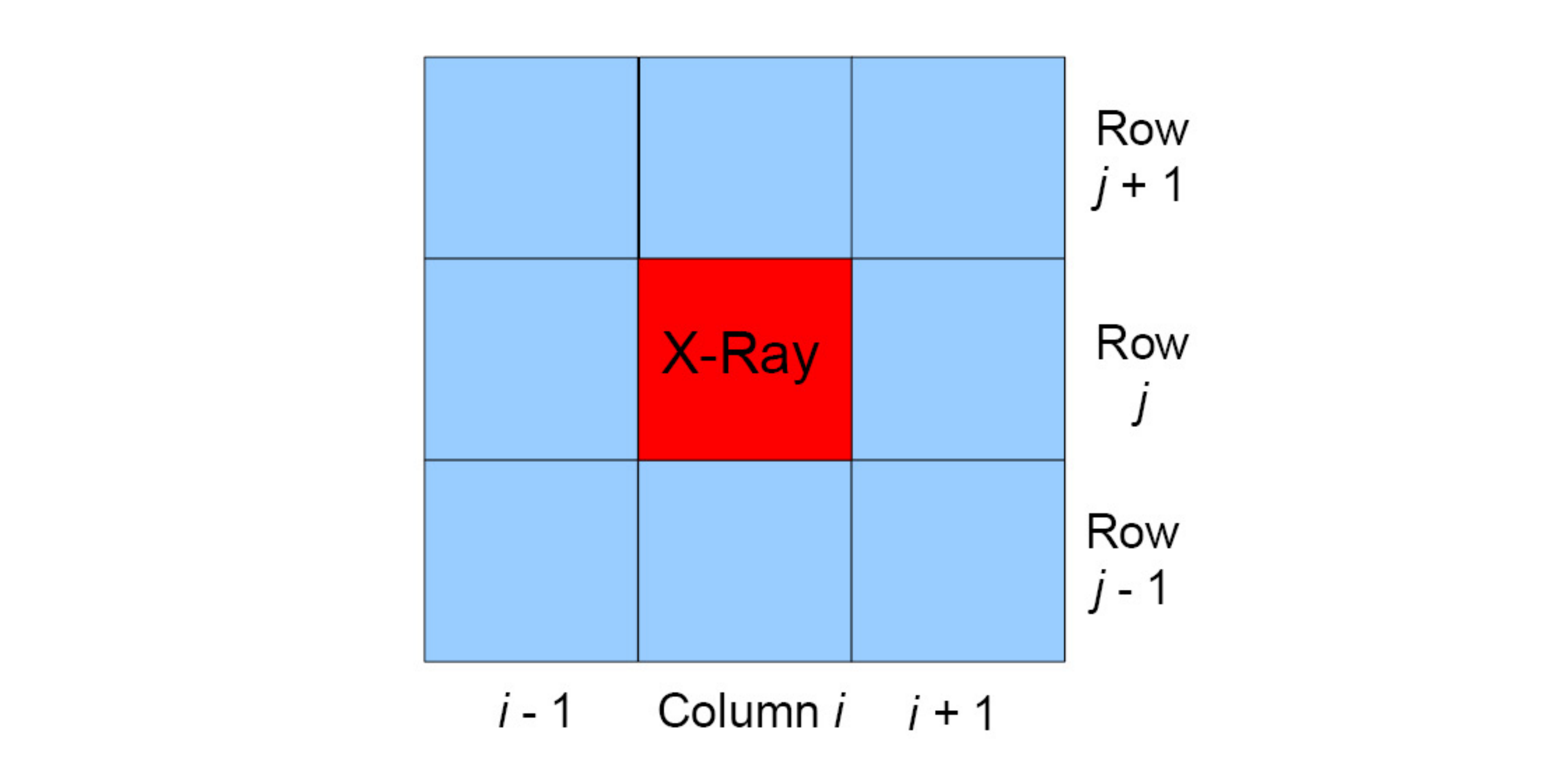}
    \caption{Illustration of the $3\times3$ cluster method. A signal charge (X-ray) is only 
accepted if no X-ray charge is present in  the  neighboring pixels.
 The X-ray charge is defined by its threshold determined after a first fit of the X-ray signal peak with a Gaussian function.}
    \label{fig:3by3_Clustering}
\end{figure}

In order to remove hits with shared charges, a 3$\times$3 cluster method is used (Fig.~\ref{fig:3by3_Clustering}). 
This method is only applied to the 3 adjacent columns (1, 2 and 3). 
Column 2 is considered, where the signal charge of a pixel is accepted only if no X-ray charge  is present in  the  neighboring pixels. 

\begin{figure}
    \centering
    \includegraphics[width=\columnwidth,clip]{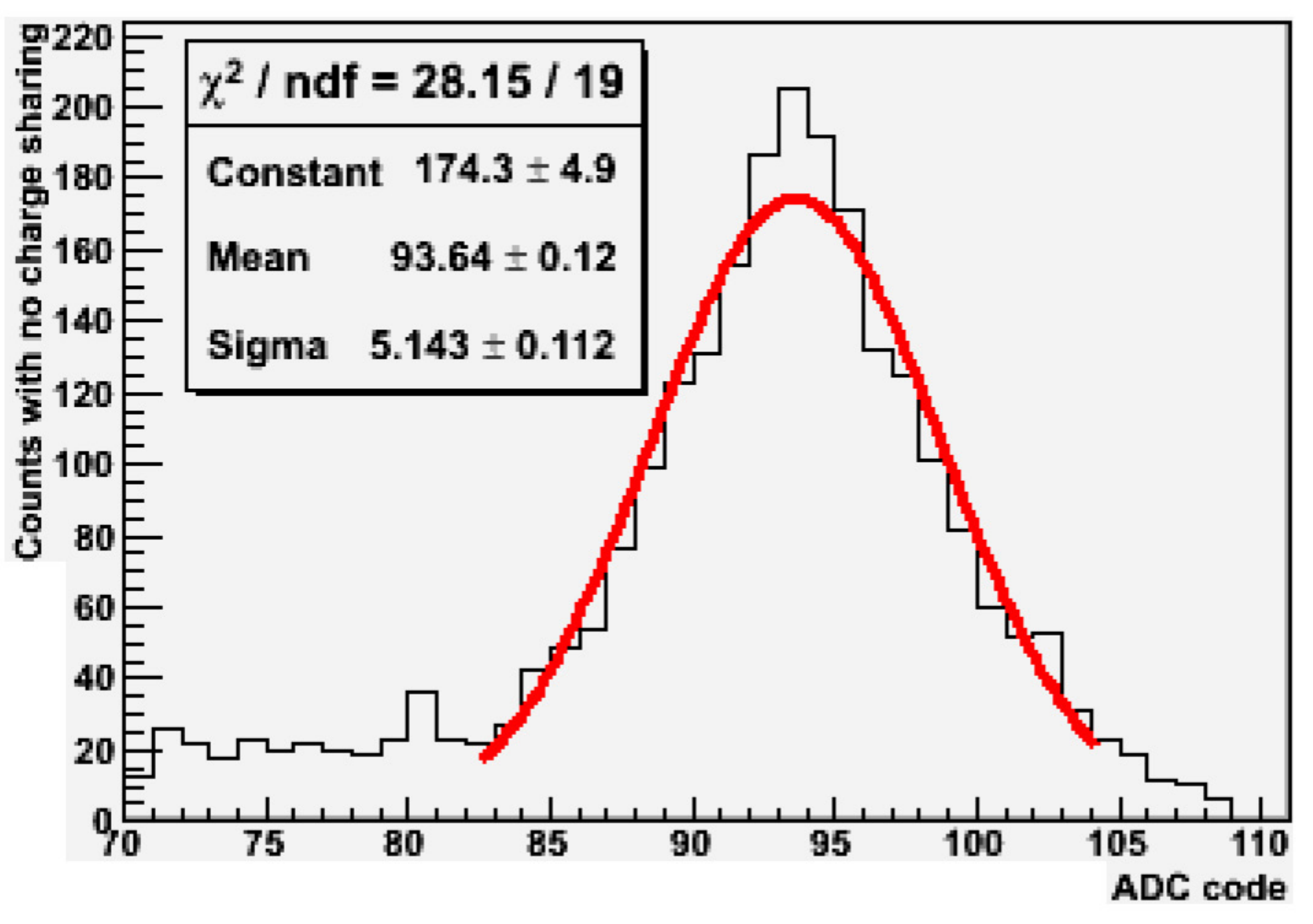}
    \caption{X-ray signal peak with a tail from some charge sharing even after 
applying the $3\times3$ cluster method. The distribution is fitted with a 
Gaussian function to determine the threshold to be used to completely remove the charge sharing.
This is a different dataset than that in Fig.~\ref{fig:Fit_NoisePeak}, 
which was obtained using a higher preamplifier gain.}
    \label{fig:Further_Removing_1}
\vspace*{4mm}
\end{figure}

\begin{figure}
    \centering
    \includegraphics[width=\columnwidth,clip]{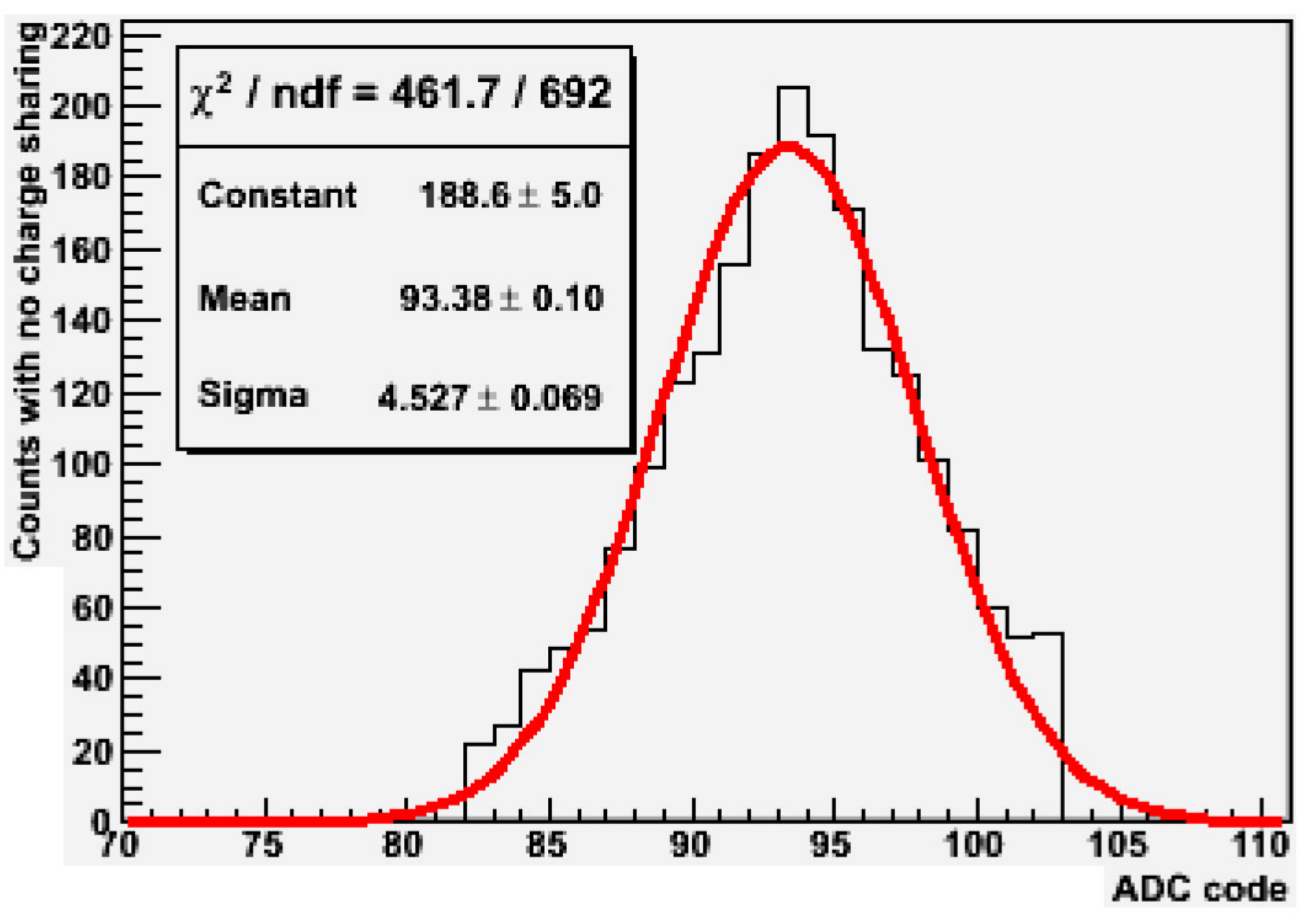}
    \caption{X-ray signal peak with no charge sharing.}
    \label{fig:Further_Removing_2}
\vspace*{5mm}
\end{figure}

The noise threshold is used to separate noise from shared X-ray charges. 
It is determined by fitting the noise peak with a Gaussian function. 
The threshold is usually set to be $5\sigma$ above the fitted peak centroid.
After removal of the  noise and reduction of the charge-sharing events, a tail 
remains for the X-ray peak as shown in Fig.~\ref{fig:Further_Removing_1}. 
In order to separate the X-ray signal region from the charge-sharing region we exclude all ADC codes
 below an X-ray threshold which is determined by fitting the X-ray peak. We usually set the threshold to be 
 $2 \sigma$ below the fitted peak centroid but with weaker X-ray peaks the fitted Gaussian 
sometimes failed to follow the real shape of the peak,
and then the limits were set by visual inspection. In this way we obtain the measured X-ray signal data in the pixels
 with full X-ray charge (Fig.~\ref{fig:Further_Removing_2}).
\section{Baseline correction for different operating temperatures and readout frequencies}
\begin{figure}
    \centering
    \includegraphics[width=\columnwidth,clip]{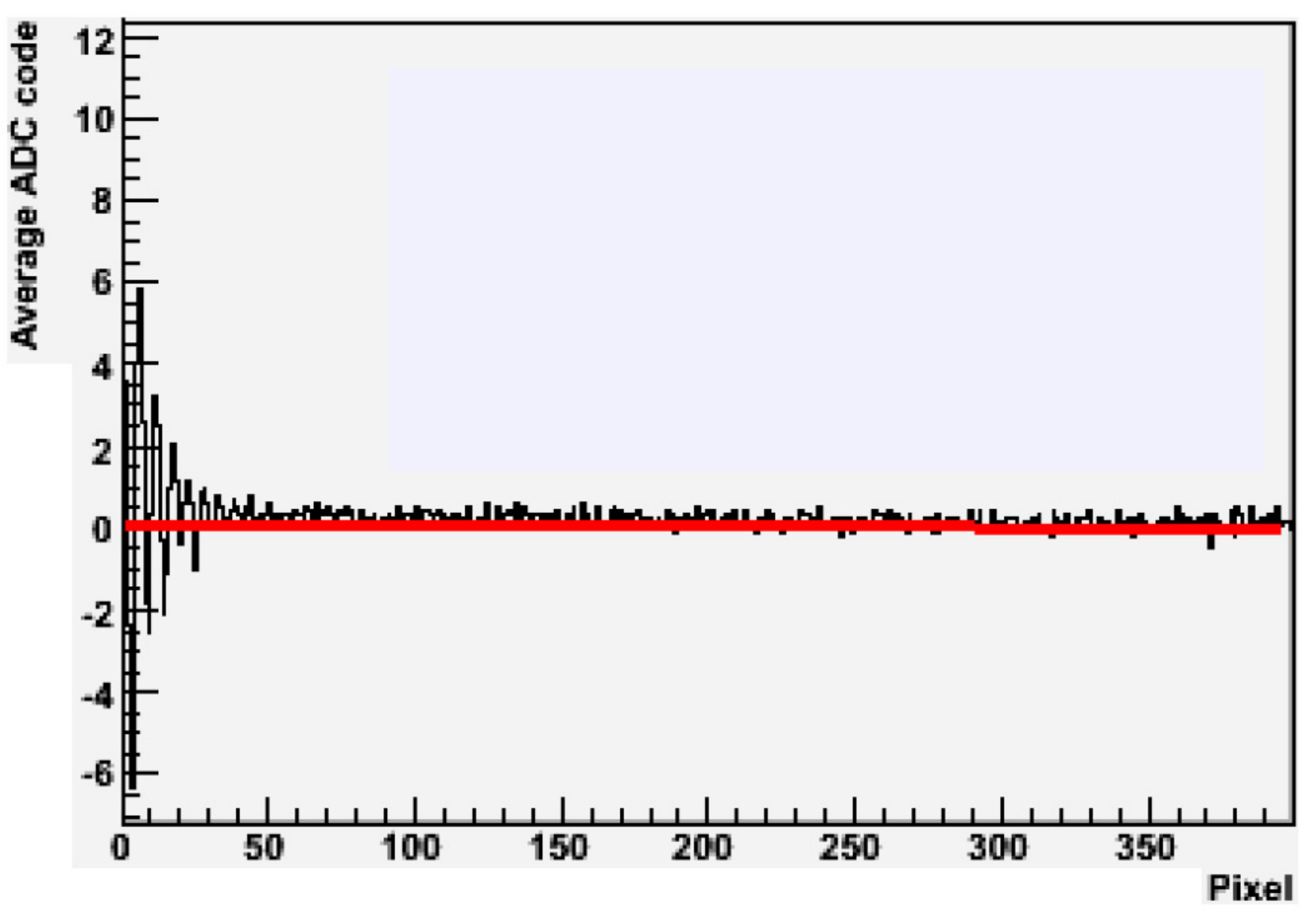}
    \caption{Distribution of average ADC code in the noise region versus the pixel number 
             at $2$~MHz and $-60~^\circ$C with an exponential fit. The noise region was delimited by a $3\sigma$  noise threshold. 
             Fit: amplitude $A=0.117 \pm 0.056$ and exponent $B=(-5.29 \pm 2.85)\times10^{-3}.$}
    \label{fig:Baseline_2_60}
\vspace*{6mm}
\end{figure}

\begin{figure}
    \centering
    \includegraphics[width=\columnwidth,clip]{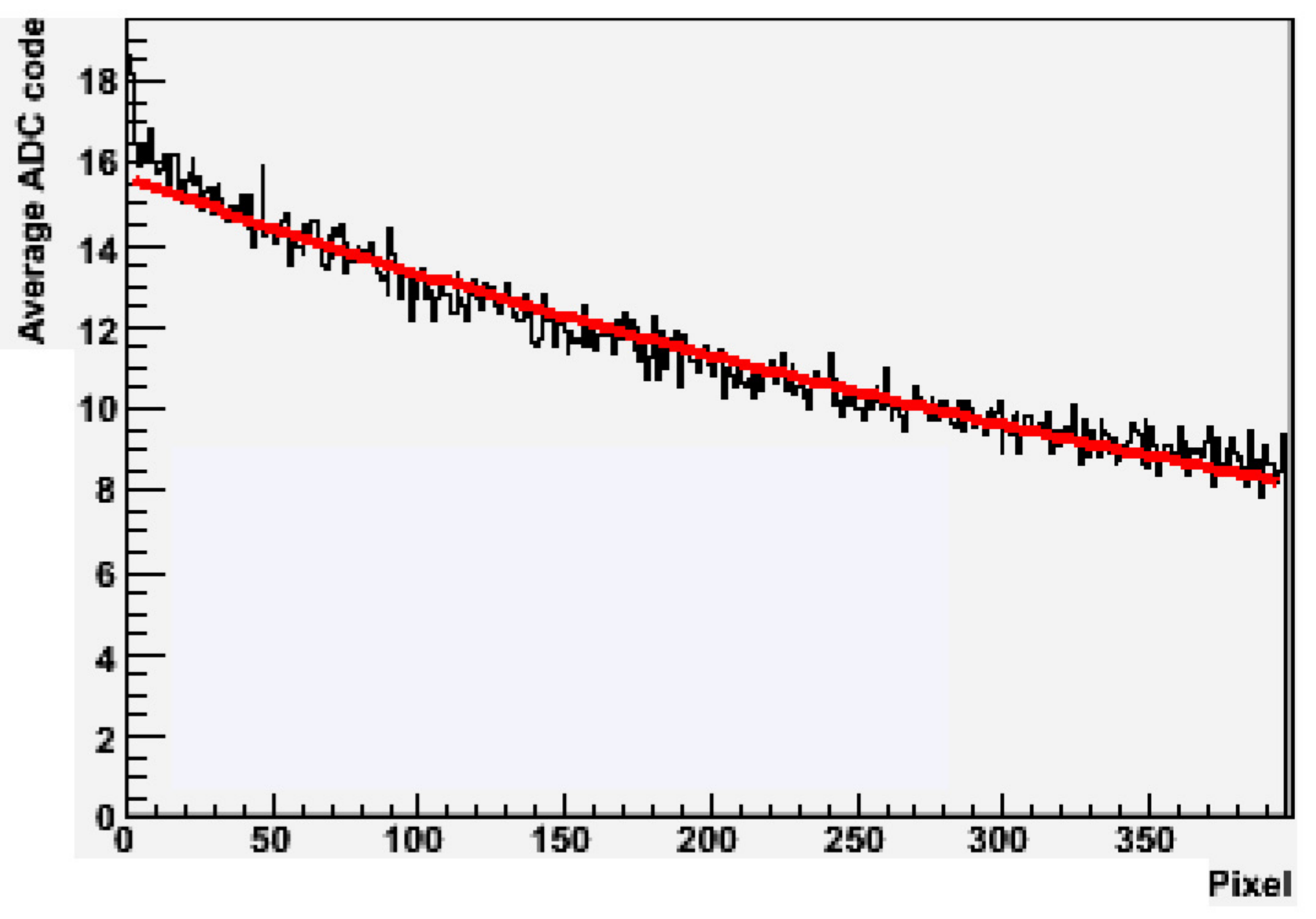}
    \caption{Distribution of average ADC code in the noise region versus the pixel number at $4$~MHz and $-6~^\circ$C 
             with an exponential fit. The noise region was delimited by a $3\sigma$  noise threshold. Fit: amplitude $A=15.64 \pm 0.44$ and exponent $B=(-16.24 \pm 1.33)\times10^{-4}.$}
    \label{fig:Baseline_4_6}
\end{figure}

\begin{figure}
    \centering
    \includegraphics[width=\columnwidth,clip]{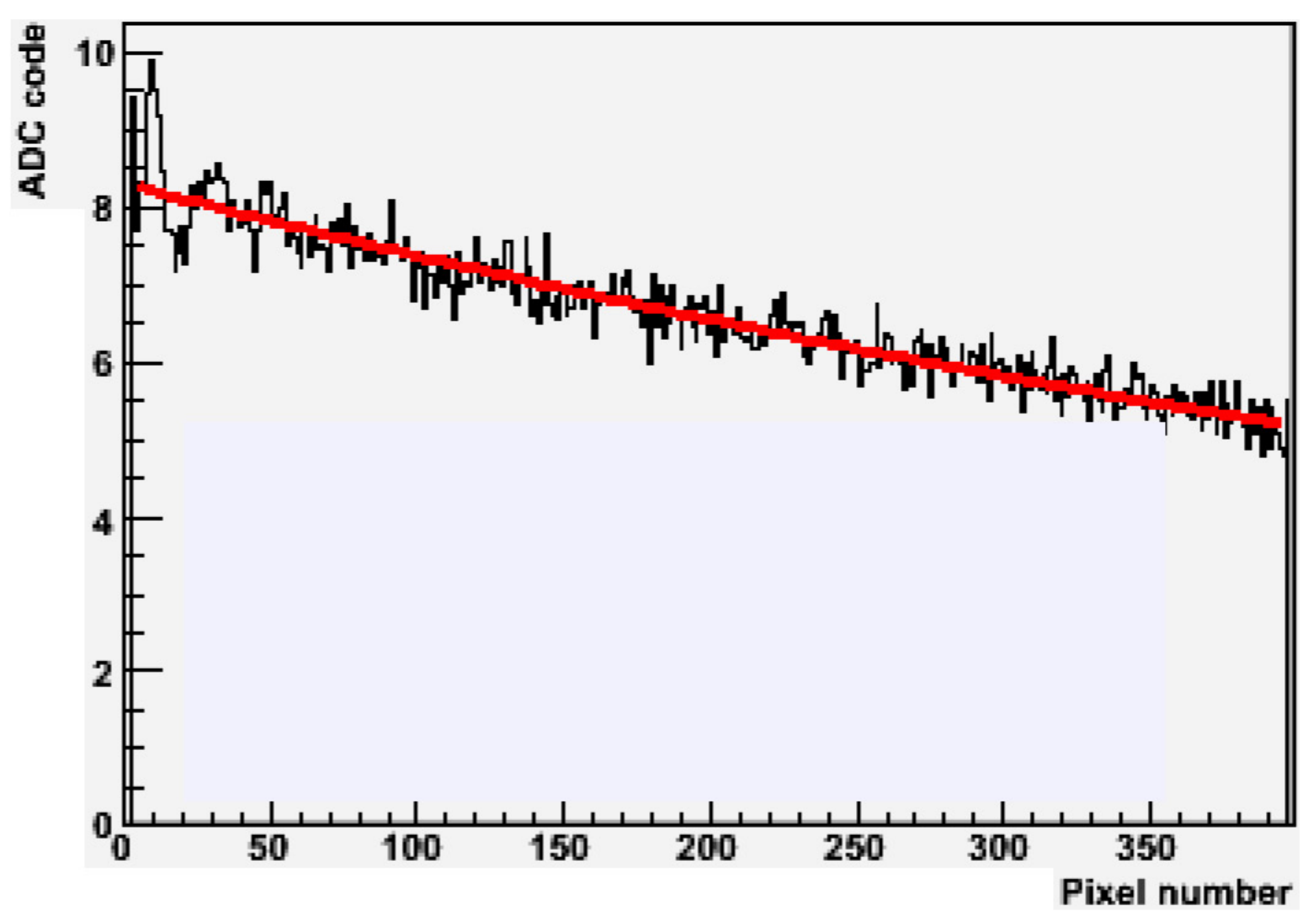}
    \caption{Distribution of average ADC code in the noise region versus the pixel number at $8$~MHz and $-17~^\circ$C 
             with an exponential fit. The noise region was delimited by a $3\sigma$  noise threshold. Fit: amplitude $A=8.31 \pm 0.31$ 
             and exponent $B=(-11.86 \pm 1.74)\times10^{-4}.$}
    \label{fig:Baseline_8_17}
\end{figure}

\begin{figure}
    \centering
    \includegraphics[width=\columnwidth,clip]{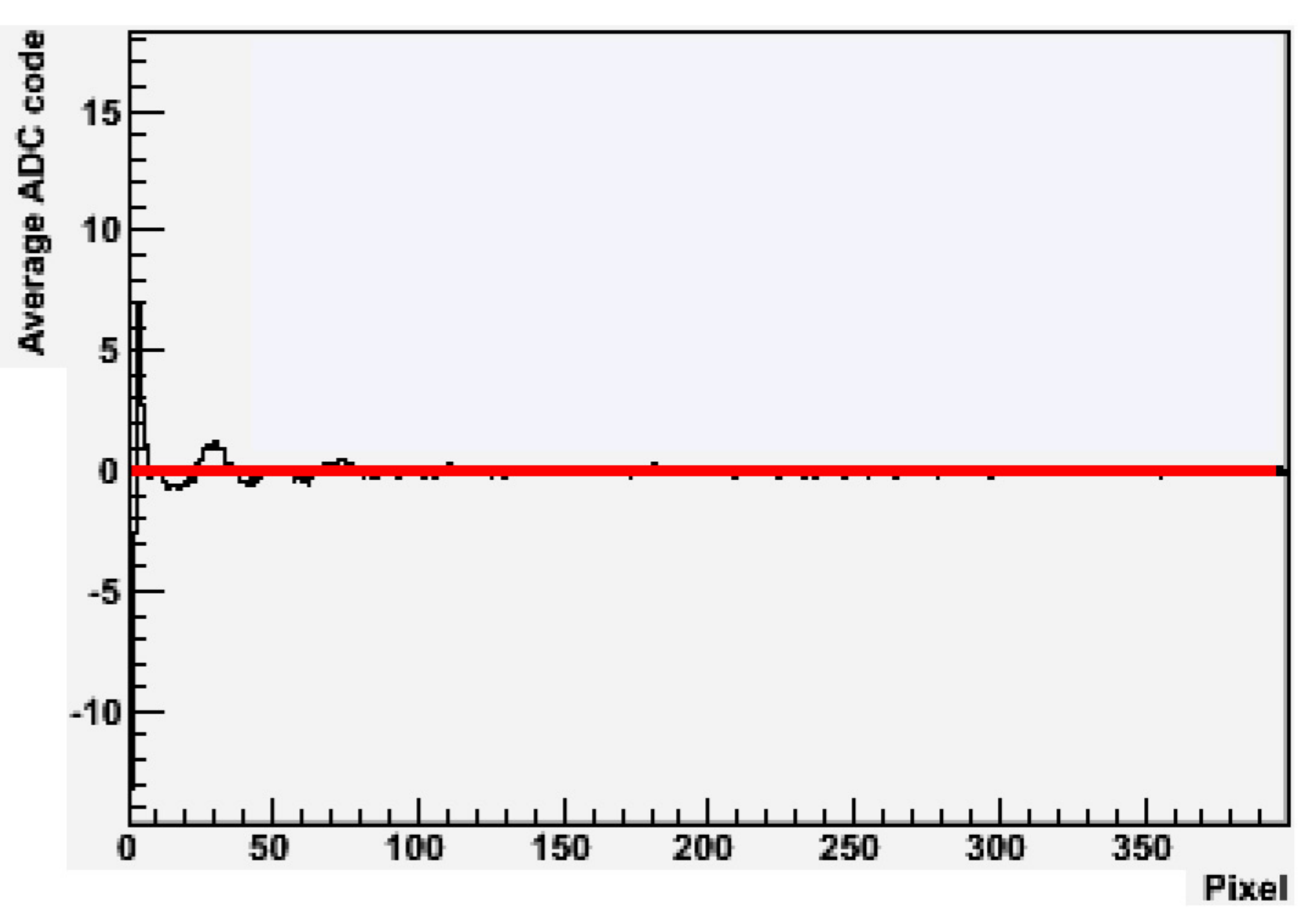}
    \caption{Distribution of average ADC code in the noise region versus the pixel number at $8$ MHz and $-109$ $^\circ$C 
             with an exponential fit. The noise region was delimited by a $3\sigma$  noise threshold. Fit: amplitude $A=(0.082 \pm 5.035)\times10^{-3}$ 
             and exponent $B=(1.073 \pm 16.853)\times10^{-2}.$}
    \label{fig:Baseline_8_109}
\vspace{-4mm}
\end{figure}

Owing to an RC coupling at the preamplifier input, the output voltage decays exponentially with a time constant
equal to $RC$ ($\tau_{\rm RC}$ is of the order of $100~\mu$s)\footnote{For a readout frequency
of $4$~MHz reading out the whole column of 400 pixels takes a time comparable to $\tau_{\rm RC}$.}.
This is expected to give rise to an exponential baseline that 
varies during the clocking sequence, adding a contribution to the 
readout signal that depends on the pixel number. The shape of 
the baseline can be determined by comparing the average output
level in the pixels to each other. For each pixel an average of the ADC codes was
 computed excluding the frames were the code
exceeded the noise threshold. The average ADC code can usually be expected 
to be an exponential function of the pixel number with the RC decay.
Figures~\ref{fig:Baseline_2_60},~\ref{fig:Baseline_4_6},~\ref{fig:Baseline_8_17} and~\ref{fig:Baseline_8_109} 
show the distributions of the average ADC codes for all frames of a measurement as a function of the  pixel number 
for different readout frequencies and operating temperatures. These distributions are fitted by an exponential 
function given by $A\exp(-Bj)$, where $A$ corresponds to the signal charge at the first pixel, 
$B$ is the slope and $j$ is the pixel number. This function will be subtracted from the X-ray data. 
Figures~\ref{fig:Baseline_2_60} and~\ref{fig:Baseline_8_109} show approximately zero baseline level, 
whereas Figs.~\ref{fig:Baseline_4_6} and~\ref{fig:Baseline_8_17} show non-zero baseline fits. 
The baseline level is much smaller at low temperatures, for example at $-60~^\circ$C (Fig.~\ref{fig:Baseline_2_60}) 
and at $-109~^\circ$C (Fig.~\ref{fig:Baseline_8_109}). 
Figures~\ref{fig:Baseline_4_6} and~\ref{fig:Baseline_8_17} show baselines for higher 
temperatures $-6~^\circ$C and $-17~^\circ$C, respectively.
The observed baselines show different frequency and temperature dependence from what is expected from the RC effect.
This indicates that other electronic effects also influence the baseline.
\section{Determination of Charge Transfer Inefficiency}
\begin{figure}
    \centering
    \includegraphics[width=\columnwidth,clip]{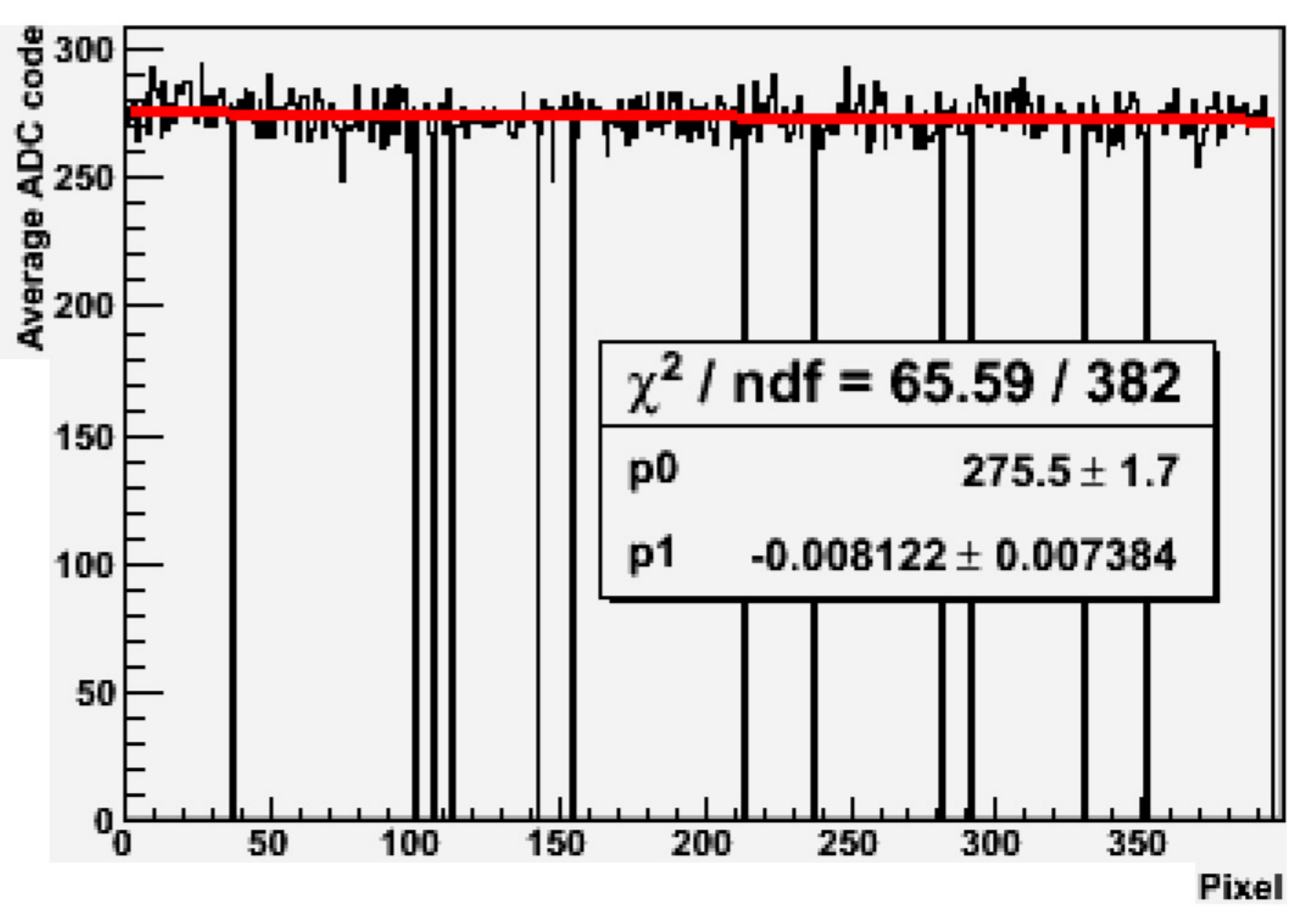}
    \caption{Fit to average ADC codes without baseline removal 
     at $2$~MHz and $-31~^\circ$C. $CTI = (2.99 \pm 2.69)\times10^{-5}$.}
    \label{fig:CTI_Fit_NoReduction_avr}
\vspace*{5mm}
\end{figure}

\begin{figure}
    \centering
    \includegraphics[width=\columnwidth,clip]{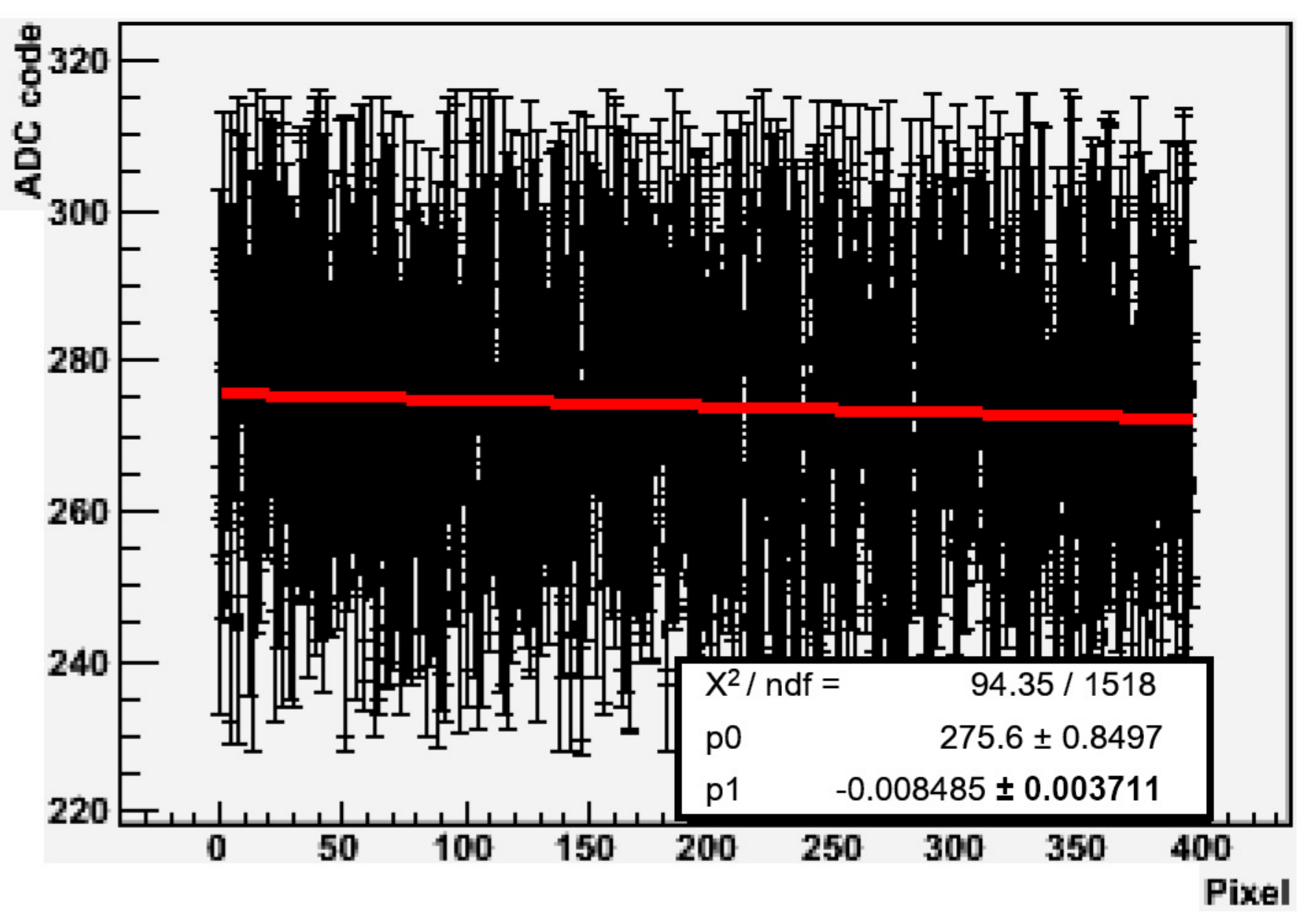}
    \caption{Fit to events (hits) without baseline removal at $2$~MHz and $-31~^\circ$C. $CTI = (3.69 \pm 1.31)\times10^{-5}$.}
    \label{fig:CTI_Fit_NoReduction_Events}
\vspace*{5mm}
\end{figure}

\begin{figure}
    \centering
    \includegraphics[width=\columnwidth,clip]{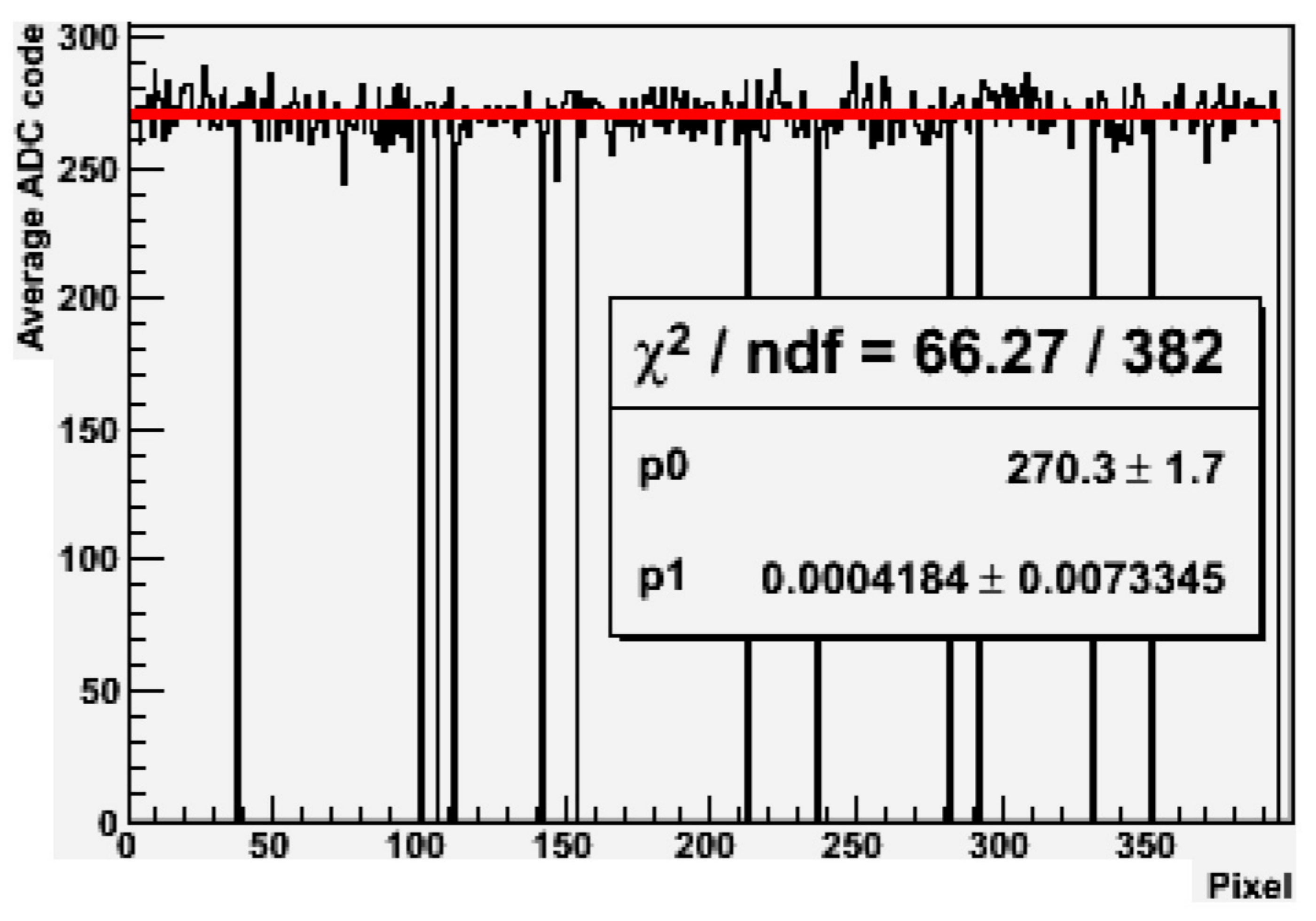}
    \caption{Fit to average ADC codes with baseline removal at $2$~MHz and $-31~^\circ$C. 
    $CTI = (-0.09 \pm 2.70)\times10^{-5}$.}
    \label{fig:CTI_Fit_Reduction_avr}
\vspace*{5mm}
\end{figure}

\begin{figure}
    \centering
    \includegraphics[width=\columnwidth,clip]{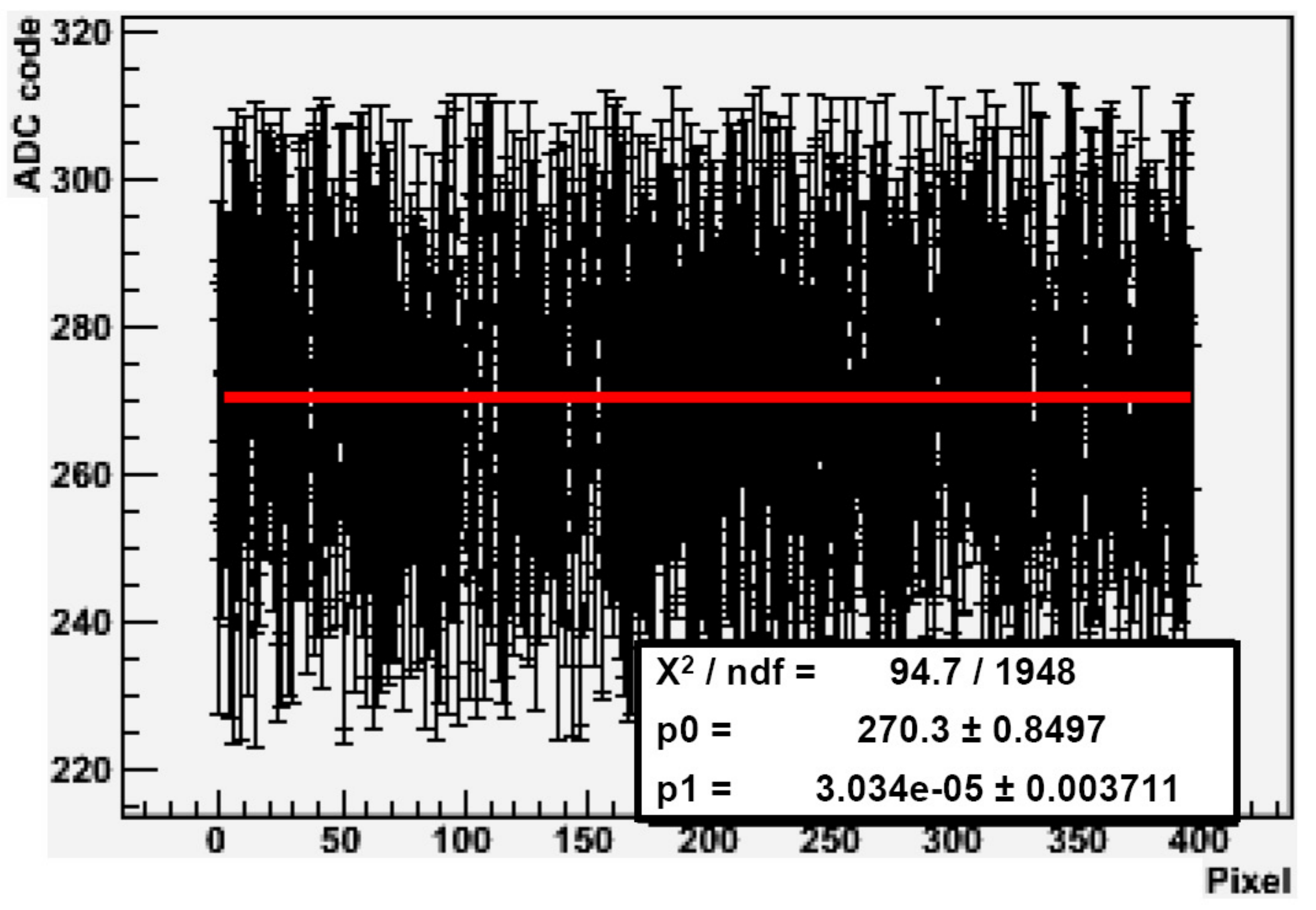}
    \caption{Fit to events (hits) with baseline removal at $2$~MHz and $-31~^\circ$C. $CTI = (0.63 \pm 1.33)\times10^{-5}$.}
    \label{fig:CTI_Fit_Reduction_Events}
\end{figure}

The charge transfer inefficiency (CTI) in one pixel is defined as the ratio of 
signal lost during transfer (captured by traps) to the initial signal charge.
 For an un-irradiated CCD we expect that the CTI value is consistent with zero within uncertainties. 
In order to determine the CTI we first make the  overclock correction and apply the $3\times3$ cluster method. 
Then, the X-ray peak was fitted with the Gaussian function to determine the X-ray threshold. 
This was used to construct the two distributions with a fit of average ADC codes as shown 
in Fig.~\ref{fig:CTI_Fit_NoReduction_avr} and with a fit of individual 
events (X-ray hits) as shown in Fig.~\ref{fig:CTI_Fit_NoReduction_Events}. 
Both distributions are fitted with the first-order polynomial function $P_0+P_1j$, where $P_0$ corresponds 
to the charge at the first pixel, $P_1$ is the slope and $j$ is the pixel number. These distributions are 
plotted without baseline removal. Figures~\ref{fig:CTI_Fit_Reduction_avr} and~\ref{fig:CTI_Fit_Reduction_Events} 
show the distributions of the averages and all events as a function of pixel number with baseline removal. 
The CTI is determined for the four cases using~$CTI=-P_1/P_0.$
\section{CTI results pre-irradiation for different readout frequencies}
\begin{figure}
    \centering
    \includegraphics[width=\columnwidth,clip]{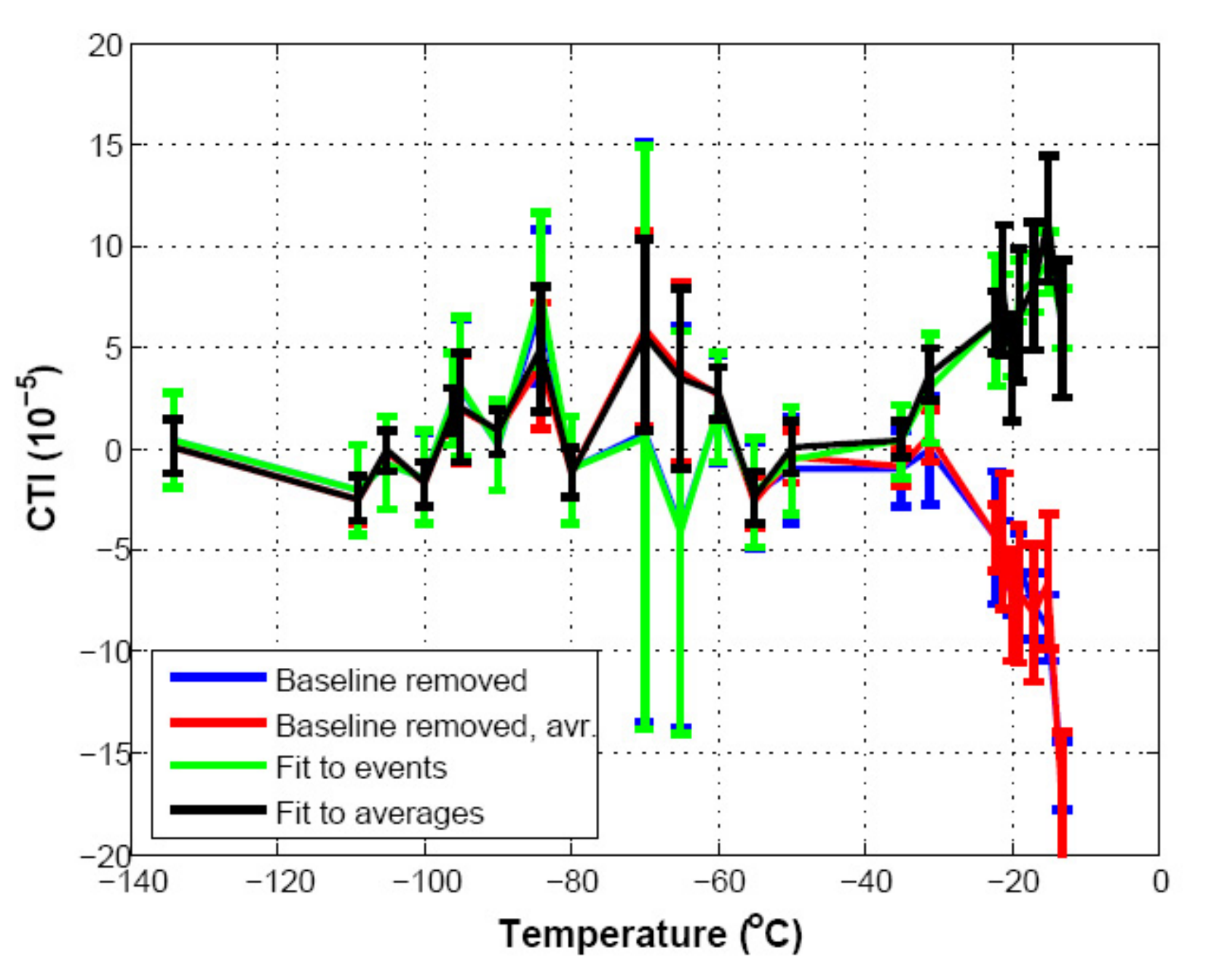}
    \caption{CTI versus temperature at $2$~MHz readout frequency. The CTI is shown resulting from fits
             to average ADC codes and to events (hits), with and without baseline removal.
             The shown error bars result from the precision of the fits.}
    \label{fig:CTI_Results_2MHz}
\end{figure}

\begin{figure}
    \centering
    \includegraphics[width=\columnwidth,clip]{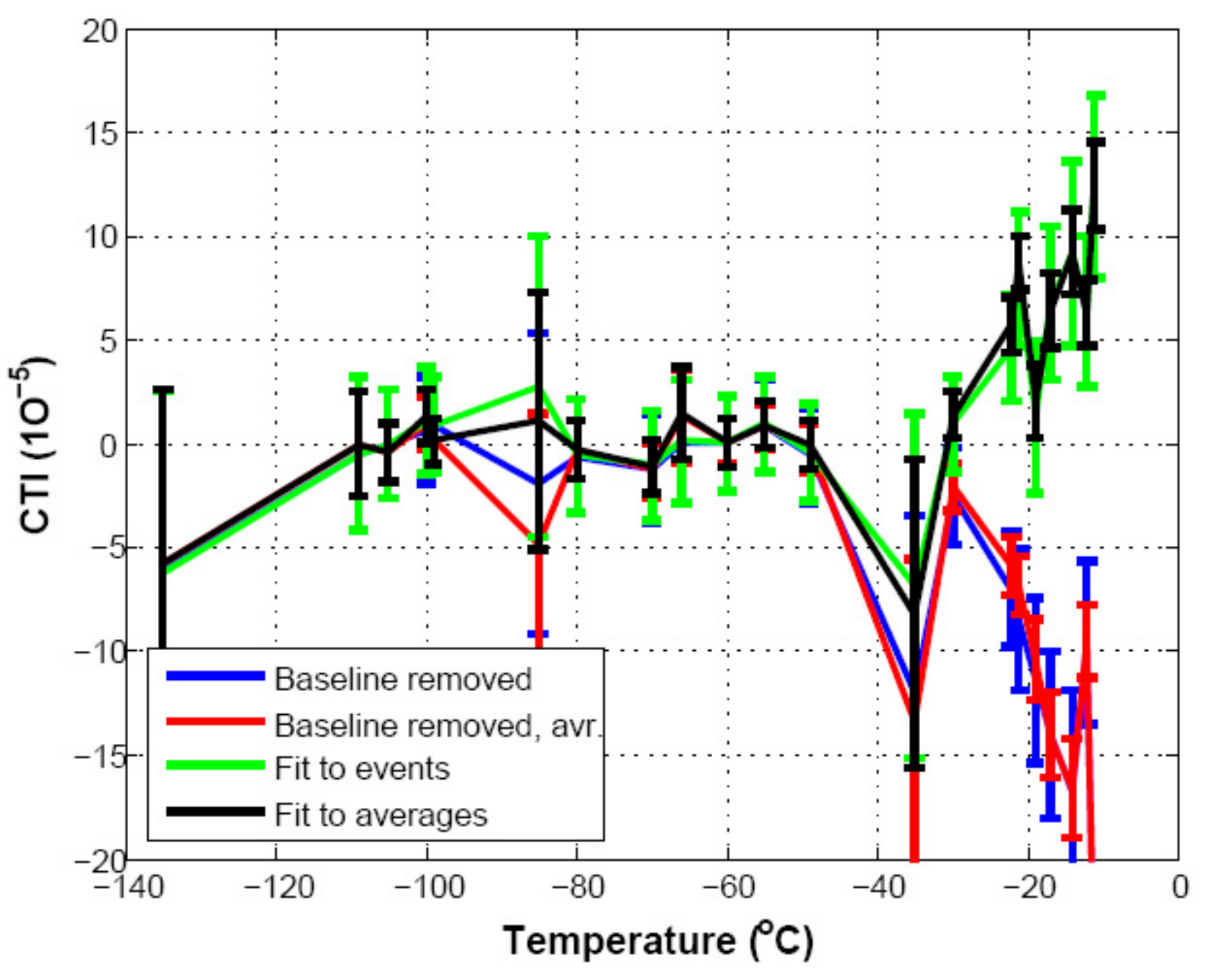}
    \caption{CTI versus temperature at $4$~MHz readout frequency. The CTI is shown resulting from fits
             to average ADC codes and to events (hits) , with and without baseline removal.
             The shown error bars result from the precision of the fits.}
    \label{fig:CTI_Results_4MHz}
\end{figure}

\begin{figure}
    \centering
    \includegraphics[width=\columnwidth,clip]{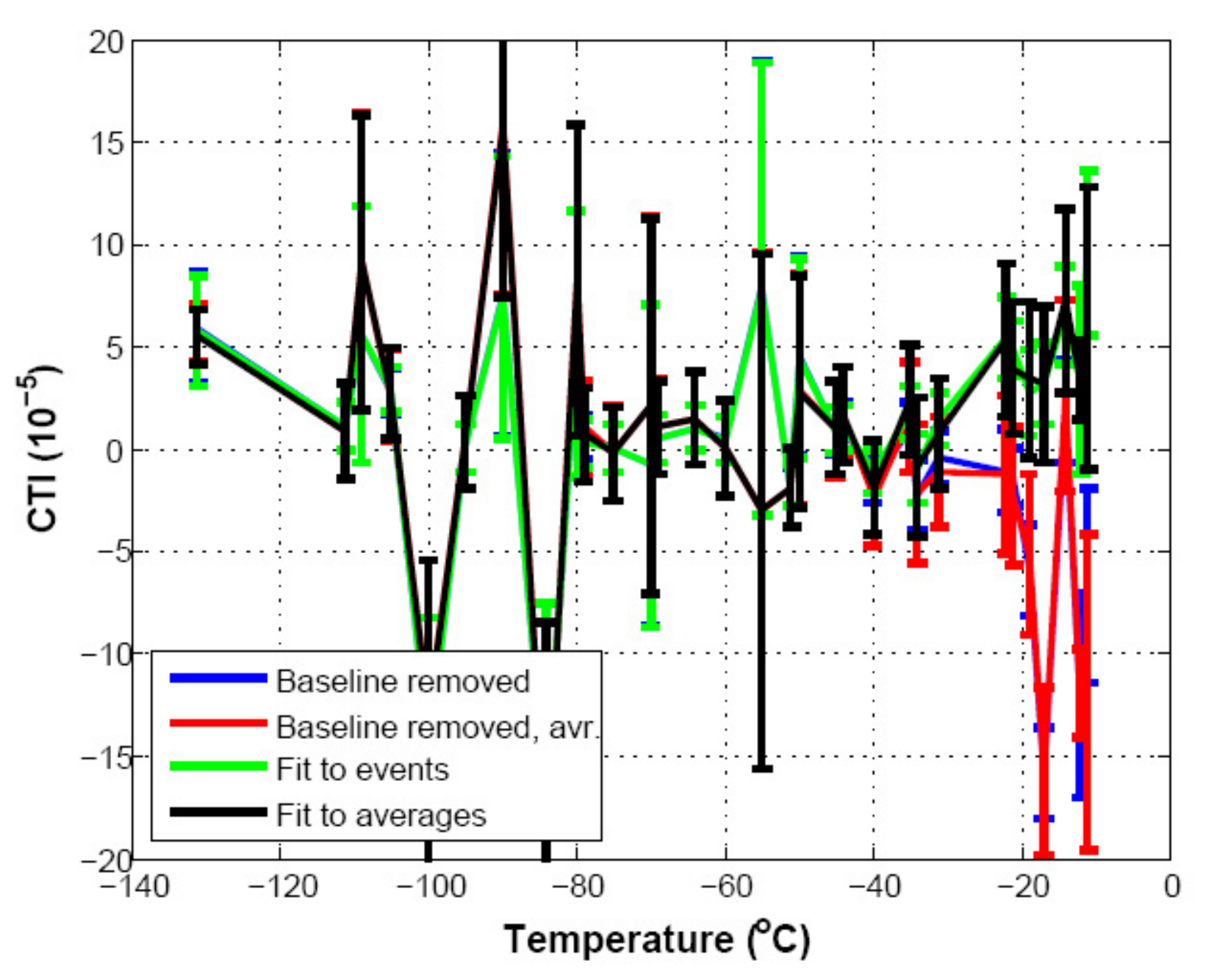}
    \caption{CTI versus temperature at $8$~MHz readout frequency. The CTI is shown resulting from fits
             to average ADC codes and to events (hits) , with and without baseline removal.
             The shown error bars result from the precision of the fits.}
    \label{fig:CTI_Results_8MHz}
\end{figure}

Figures~\ref{fig:CTI_Results_2MHz},~\ref{fig:CTI_Results_4MHz} and~\ref{fig:CTI_Results_8MHz} show the CTI 
values as a function of temperature for an un-irradiated CPC1 at different readout frequencies. 
The CTI has been calculated using two methods, a linear fit of averages versus pixel number and using a 
linear fit of events (X-ray hits) versus pixel number. As it is expected for an un-irradiated CCD, the CTI is small 
because the density of traps is small. In the temperature region below $-25~^\circ$C no frequency dependence of the
CTI measurements is observed. One may note that in the region of rather high 
temperatures (above $-25~^\circ$C) the effect of baseline removal is large and the determined CTI values
indicate that the baseline effect is overestimated. The effect of the baseline removal decreases with increasing frequency
as the readout clocking sequence becomes shorter compared to $\tau_{\rm RC}$ (as visible 
in Fig.~\ref{fig:CTI_Results_8MHz} for $8$~MHz readout frequency).
\section{Conclusions and Outlook}
An un-irradiated CPCCD is operated in a range of temperatures from $-10$ $^\circ$C to $-136$ $^\circ$C 
(liquid nitrogen cooling)  with different readout frequencies $2$, $4$ and $8$~MHz. 
The spectrum of a $^{55}$Fe source  is measured with this device. 
The CTI is analysed for different readout frequencies and operating temperatures. 
A clear X-ray signal is extracted by identifying isolated hits ($3\times3$ method). 
The baseline is subtracted. The CTI value is small and compatible with zero as can be expected for an un-irradiated CPCCD. 
Further CTI measurements with a CPCCD after irradiation and refinement of the analysis method are planned.
\vspace*{-1mm}
\section*{Acknowledgements}
\vspace*{-1mm}
We would like to thank Salim Aoulmit, Alex Chilingarov and Lakhdar Dehimi for discussions and comments on the manuscript.
This work is supported by the Science and Technology Facilities Council (STFC) and Lancaster University. 
KB wishes to thank the Algerian Government for financial support and Lancaster University for their hospitality. 
AS would like to thank the Faculty of Science and Technology at Lancaster University for financial support 
and the organizers of the IEEE`08 conference for their hospitality.
\vspace*{-1mm}
\bibliographystyle{mcite}

\begin{thebibliography}{99}
\vspace*{-1mm}
\bibitem{Damerell:1997vv} C.J.S. Damerell, ``Radiation damage in CCDs used as particle detectors'', ICFA Instrum. Bull. 14 (1997) 1.
\bibitem{Stefanov} K. Stefanov, PhD thesis, Saga University (Japan), ``Radiation damage effects in CCD sensors for tracking applications in high energy physics'', 2001, and references therein;
K. Stefanov et al., ``Electron and neutron radiation damage effects on a two phase CCD'', IEEE Trans. Nucl. Sci. 47 (2000) 1280.
\bibitem{LCFI_web} LCFI collaboration homepage: http://hepwww.rl.ac.uk/lcfi/.
\bibitem{Worm} S.D. Worm, ``Recent CCD development for the vertex detector of the ILC -- Including ISIS (In-situ Storage Image Sensors)'', in 10th Topical Seminar on Innovative Particle and Radiation Detectors (IPRD06), Siena, Italy, October 1-5, 2006.
\bibitem{Greenshaw} T.J. Greenshaw, ``Column Parallel CCDs and in-situ storage image sensors for the vertex detector of the international linear collider'', in 2006 Nuclear Science Symposium, San Diego, USA, October 29-November 4, 2006.
\bibitem{Marconi} M.S. Robbins ``The Radiation Damage Performance of Marconi CCDs'', Marconi Technical Note S\&C 906/424 2000 (unpublished).
\bibitem{Brau2000} J.E. Brau and N.B. Sinev, ``Operation of a CCD particle detector in the presence of bulk neutron damage'', IEEE Trans. Nucl. Sci. 47 (2000) 1898.
\newpage
\bibitem{Brau2005}J.E.~Brau, O.~Igonkina, C.T.~Potter and N.B.~Sinev, ``Investigation of radiation damage effects in neutron irradiated CCD'', Nucl. Instr. and Meth. A549 (2005) 1173.
\bibitem{Mohsen} A.M. Mohsen and M.F. Tompsett, ``The effect of bulk traps on the performance of bulk channel charge-coupled devices'', IEEE Trans. Electron Dev. ED21, 11 (1974) 701.
\bibitem{Hopkins} I.H. Hopkins, G. Hopkinson and B. Johlander, ``Proton-induced charge transfer degradation in 
CCD's for near-room temperature applications'', IEEE Trans. Nucl. Sci. 41 (1994) 1984.
\bibitem{Hardy} T. Hardy, R. Murowinski and M.J. Deen, ``Charge transfer efficiency in proton damaged CCD's'', IEEE Trans. Nucl. Sci. 45 (1998) 154.
\bibitem{Sopczak2005} A.~Sopczak, ``LCFI Charge Transfer Inefficiency Studies for CCD Vertex Detectors'', IEEE 2005 Nuclear Science Symposium, San Juan, USA. Proc. IEEE Nuclear Science Symposium Conference Record N37-7 (2005) 1494.
\bibitem{Sopczak2006} A.~Sopczak, ``Radiation Hardness of CCD Vertex Detectors for the ILC'', IEEE 2006 Nuclear Science Symposium, San Diego, USA. October 29-November 4, 2006. Proc. IEEE Nuclear Science Conference Record N14-215 (2006) 576.
\bibitem{Sopczak2007} A. Sopczak et al., ``Simulations of the Temperature Dependence of the Charge Transfer Inefficiency in a High-Speed CCD'', IEEE Trans. Nucl. Sci. 54 (2007) 1429, and references therein.
\bibitem{IEEE_Honolulu} A. Sopczak et al., ``Radiation Hardness Studies in a CCD with High-Speed Column Parallel Readout'', 
Proc. 2007 IEEE Nuclear Science Symposium, October 27-November 3, 2007, Honolulu, USA. Proc. IEEE Nuclear Science Conference Record N48-2 (2007) 2278.
\bibitem{Sopczak2008} A. Sopczak et al., ``Radiation Hardness Studies in a CCD with High-speed Column Parallel Readout'', JINST 3 (2008) 5007.
\bibitem{Walker} J.W. Walker and C.T. Sah, `` Properties of 1.0-MeV-electron-irradiated defect centers in silicon'', Phys. Rev. B7 (1972) 4587.
\bibitem{Saks} N.S.~Saks, ``Investigation of Bulk Electron Traps Created by Fast Neutron Irradiation in a Buried N-Channel CCD'', IEEE Trans. Nucl. Sci. 24 (1977) 2153.
\bibitem{Srour} J.R.~Srour, R.A.~Hartmann and S.~Othmer, ``Transient and Permanent Effects of Neutron Bombardment on a Commercially Available N-Buried-Channel CCD'', IEEE Trans. Nucl. Sci. 27 (1980) 1402.
\bibitem{Root_Web} ROOT homepage: http://root.cern.ch/.
\vspace*{-0.2cm}
\end{thebibliography}

\end{document}